# The Capacity of MIMO Channels with Per-Antenna Power Constraint

Mai Vu

Department of Electrical and Computer Engineering, McGill University,

Montreal, H3A2A7

Email: mai.h.vu@mcgill.ca

### Abstract

We establish the optimal input signaling and the capacity of MIMO channels under per-antenna power constraint. While admitting a linear eigenbeam structure, the optimal input is no longer diagonalizable by the channel right singular vectors as with sum power constraint. We formulate the capacity optimization as an SDP problem and solve in closed-form the optimal input covariance as a function of the dual variable. We then design an efficient algorithm to find this optimal input signaling for all channel sizes. The proposed algorithm allows for straightforward implementation in practical systems in real time. Simulation results show that with equal constraint per antenna, capacity with per-antenna power can be close to capacity with sum power, but as the constraint becomes more skew, the two capacities diverge. Forcing input eigenbeams to match the channel right singular vectors achieves no improvement over independent signaling and can even be detrimental to capacity.

## I. INTRODUCTION

The capacity of a MIMO wireless channel depends on transmit power constraints and the availability of channel state information (CSI) at the transmitter and receiver. With sum power constraint across all transmit antennas, the capacity and optimal signaling are well established. For channel state known at both the transmitter and receiver, the capacity can be obtained by performing singular value decomposition of the channel matrix and water-filling power allocation on the channel eigenvalues [1]. For Rayleigh fading channels with state known only at the receiver, the ergodic capacity is obtained by sending independent signals with equal power from all transmit antennas [2].

Under the per-antenna power constraint, the MIMO capacity is less well understood. However, this per-antenna power constraint is more realistic in practice than sum power because each antenna is usually connected to a separate power amplifier on the individual RF chain. Each power amplifier has its own dynamic range, hence the transmitter may not be able to allocate power arbitrarily among its antennas. Another appealing scenario for per-antenna constraint is in a distributed MIMO system, which has transmit antennas located at different physical nodes that cannot share power with each other. Thus understanding the capacity and optimal signaling schemes under per-antenna power is useful.

The per-antenna power constraint has been investigated under different settings. In [3], the problem of a multi-user downlink channel is considered with per-antenna power constraint. For downlink broadcast, the capacity optimization problem is non-convex. It was argued that linear processing at both the transmitter (by multi-mode beamforming) and the receiver (by MMSE receive beamforming with successive interference cancellation) can achieve the capacity region. Using uplink-downlink duality, the boundary points of the capacity region for the downlink channel with per-antenna constraint can be found by solving a dual uplink problem, which maximizes a weighted sum rate for the uplink channel with sum power constraint across the users and an uncertain noise. Even though the dual uplink problem is convex, no efficient algorithms yet exist for solving it.

Similar uplink-downlink duality holds with sum power constraint, in which case, the uplink capacity can be solved using the efficient and distributed iterative water-filling algorithm [4]. This algorithm is based on the closed-form solution for capacity of a *single-user* MIMO channel with sum power constraint. Yet again for per-antenna power constraint, no capacity solutions so far exist even for single-user systems. The lack of such a single-user solution implies full and centralized computation required to solve multi-user problems, including the convex problem of uplink capacity.

In a previous paper [5], we have established the capacity of a single-user MISO channel (with single receive antenna) under the per-antenna constraint. The MISO channel admits a closed-form solution for the optimal input signaling as single-mode beamforming with only the beam phases being matched to the channel while the amplitudes are determined by the power constraint.

In this paper, we solve the single-user MIMO capacity with per-antenna. With perfect CSI at both transmitter and receiver, it can be shown that channel eigen-beamforming is no longer optimal, as was the case with sum power constraint. We formulate the capacity optimization problem in the SDP framework and analyze the optimality conditions. As the optimal input is Gaussian, we establish in closed-form its covariance matrix as a function of the dual variable. We then propose a simple, iterative algorithm to find this optimal input covariance and the capacity. Without CSI at the transmitter (CSIT) in Rayleigh fading channels, a simple analysis shows that the optimal signaling scheme is to send independent Gaussian signals from all transmit antennas.

This paper consists of 9 sections, with problem setups and preliminaries in Sections II and III. Sections IV–VI discuss the case of perfect CSIT with detailed solutions and algorithms. Section VII discusses the case of Rayleigh fading channel without CSIT. Numerical examples and analysis are provided in Section VIII before conclusion in Section IX.

For notation, we use bold face lower-case letters for vectors, bold face capital letters for matrices, $(.)^T$ for transpose, $(.)^*$ for conjugate, $(.)^\dagger$ for conjugate transpose, $\succcurlyeq$ and $\preccurlyeq$ for matrix inequalities (positive semi-definite relation), tr$(.)$ for trace, diag$\{.\}$ or diag$(.)$ for forming a diagonal matrix with the specified elements or from the diagonal values of the specified matrix.

## II. Channel Model and Power Constraints

### A. Channel model

Consider a frequency-flat multiple-input multiple-output (MIMO) channel with $n$ transmit and $m$ receive antennas. The channel between each transmit-receive pair is a complex, multiplicative factor $h_{ij}$. Denote the channel coefficient matrix as $\mathbf{H}$ of size $m \times n$, and the transmit signal vector as $\mathbf{x} = [x_1 \dots x_n]^T$. Then the received signal vector of length $m$ can be written as

$$\mathbf{y} = \mathbf{H}\mathbf{x} + \mathbf{z} \tag{1}$$

where $\mathbf{z} \sim \mathcal{CN}(0, \mathbf{I})$ is a vector of additive white circularly complex Gaussian noise. Here we have normalized the noise power at all receivers, which can be done by absorbing the actual noise power into the transmit power constraint.

Assume perfect CSI at the receiver, we will consider two cases of CSI at the transmitter: perfect CSIT, and no CSIT. Perfect CSIT with $\mathbf{H}$ known also to the transmitter can be applied to slow fading channels, in which channel tracking (by any mean of reciprocity or feedback) is possible, or to non-fading channels such as those in digital subscriber lines. No CSIT is applicable in fast-fading wireless channels, in which the transmitter only knows the channel distribution. For ergodic capacity, we model $\mathbf{H}$ as a random, circular complex Gaussian matrix with zero mean and covariance $\mathbf{I}_{m \times n}$.

The capacity of MIMO channel (1) depends on the power constraint on input signal vector $\mathbf{x}$. In all cases, because of the Gaussian noise and known channel state at the receiver, the optimal input signal is Gaussian with zero mean [2]. Let $\mathbf{Q} = E[\mathbf{x}\mathbf{x}^\dagger]$ be the covariance of this Gaussian input, then the achievable transmission rate for a specific channel is

$$r = \log \det \left( \mathbf{I}_m + \mathbf{H}\mathbf{Q}\mathbf{H}^\dagger \right). \tag{2}$$

The remaining question is to establish the optimal $\mathbf{Q}$ that maximizes this rate or its expected value, hence achieves the capacity, according to the CSIT condition and a given power constraint.

### B. Power constraints

The MIMO capacity is often studied with sum power constraint across all antennas. In this paper, we consider a more realistic per-antenna power constraint. For comparison, we also include independent multiple-access power constraint. We elaborate on each power constraint below.

*1) Sum power constraint:* With sum power constraint, the total transmit power from all $n$ antennas is $P$, but this power can be shared or allocated arbitrarily among the transmit antennas. This constraint translates to a condition on the input covariance as

$$\text{tr}(\mathbf{Q}) \leq P. \tag{3}$$

This constraint allows complete cooperation among the transmit antennas.

*2) Independent multiple-access power constraint:* In this case, each transmit antenna has its own power budget and acts independently. This constraint can model the case of distributed transmit antennas, such as on different wireless nodes scattered in a field, without explicit cooperation among them. Let $P_i$ be the power constraint on antenna $i$, then the multiple-access constraint is equivalent to having a diagonal input covariance $\mathbf{Q} = \text{diag}\{P_i\}$. Denote $\mathbf{P} = \text{diag}\{P_i\}$, where $\text{tr}(\mathbf{P}) = P$ in relation to (3), then the multiple-access power constraint can also be expressed as

$$\mathbf{Q} \preccurlyeq \mathbf{P}. \tag{4}$$

Writing this constraint in the above semi-definite form is convenient for analyzing the capacity optimization problem later, but it does not alter the solution since to achieve the capacity, the power constraint must be met with equality and hence $\mathbf{Q} = \mathbf{P}$.

*3) Per-antenna power constraint:* Here each antenna also has a separate transmit power budget of $P_i$ $(i = 1, \ldots, n)$ but can cooperate with each other in terms of signaling. Such a channel can model a physically centralized MIMO system, in which the per-antenna constraint comes from the separate RF chain at each antenna. This channel can also model a distributed (but cooperative) MIMO system, in which each transmit antenna belongs to a sensor or ad hoc node distributed in a network. The distributed nodes have no ability to share or allocate power among themselves. The per-antenna constraint is equivalent to having the input covariance matrix $\mathbf{Q}$ with fixed diagonal values $Q_{ii} = P_i$. But different from the multiple-access constraint, the off-diagonal values of $\mathbf{Q}$ can be non-zero. Denote $\mathbf{e}_i = [0 \ldots 1 \ldots 0]^T$ as a vector with the $i^{\text{th}}$ element equal to 1 and the rest are 0. Then per-antenna constraint can also be written as

$$\mathbf{e}_i^T \mathbf{Q} \mathbf{e}_i \leq P_i, \quad i = 1 \ldots n. \tag{5}$$

This is a set of linear constraints on $\mathbf{Q}$. Again expressing the constraint as an inequality instead of equality does not alter the optimal solution. It should be stressed that constraints on the diagonal values of $\mathbf{Q}$ are not the same as constraints on the eigenvalues of $\mathbf{Q}$.

## III. Eigenbeam structure

The eigen-decomposition of $\mathbf{Q}$ is often associated with multimode beamforming. The eigenvectors of $\mathbf{Q}$ provide the beam directions, whereas the eigenvalues provide the power allocation on these beams. Next, we will review known result for capacity with sum power, which admits nicely separate solutions for the beam directions and power allocation, by performing channel singular value decomposition. We will then discuss the beam directions for per-antenna power and show that channel singular vectors are no longer applicable.

### A. Review of capacities with sum and multiple access power

With sum power constraint (3), the optimal solution is found by the well-known water-filling algorithm [1]. Perform the singular value decomposition of the channel as $\mathbf{H} = \mathbf{U_H} \mathbf{\Lambda_H} \mathbf{V_H^\dagger}$, which provides well-defined singular vectors for all the non-zero singular values. Then the capacity-optimal covariance matrix $\mathbf{Q}^\star$ has the eigenvalue decomposition as $\mathbf{Q}^\star = \mathbf{V_H^\dagger} \mathbf{\Lambda_Q} \mathbf{V_H}$, where the eigenvalues $\lambda_{\mathbf{Q},i}$ are obtained through water-filling as

$$\lambda_{\mathbf{Q},i} = \left( \mu - \lambda_{\mathbf{H},i}^{-1} \right)^+, \quad i = 1 \ldots n \tag{6}$$

for the $\lambda_{\mathbf{H},i} \neq 0$ (if $\lambda_{\mathbf{H},j} = 0$, the corresponding $\lambda_{\mathbf{Q},j} = 0$). Here $\mu$ is the water level chosen such that $\mathrm{tr}(\mathbf{Q}^\star) = P$.

Thus with sum power, the optimal solution for $\mathbf{Q}$ is diagonalizable by the channel right singular vectors. The optimal signaling is multi-mode beamforming with the beam directions specified by the channel right singular vectors (associated with the non-zero singular values) and the beam power allocation obtained by water-filling.

With multiple-access constraint (4), the obvious solution is $\mathbf{Q} = \mathbf{P}$. In this case the optimal signaling is sending independent signals from different antennas, each with the constrained power.

### B. Forced beam directions with per-antenna power

Because the channel right singular vectors provide a simple yet optimal result for the beam directions with sum power, we may be tempting to use the same beam directions for per-antenna power. But a simple analysis can show that these beam directions may not be feasible, let alone being optimal. That is, we may not always be able to find a $\mathbf{Q}$ with eigenvectors given by $\mathbf{V_H}$ that satisfies the per-antenna power constraint.

Indeed, let $\mathbf{V_H}$ be the eigenvectors of $\mathbf{Q}$ for per-antenna power, then we can write $\mathbf{Q} = \sum_{i=1}^n \lambda_{Q,i} \mathbf{v}_i \mathbf{v}_i^\dagger$, where $\mathbf{v}_i$ are columns of $\mathbf{V_H}$, and $\lambda_{Q,i}$ are the eigenvalues of $\mathbf{Q}$. We now only need to find $\lambda_{Q,i}$ to satisfy the per-antenna power $Q_{jj} = P_j$. This condition translates to

$$\sum_{i=1}^n \lambda_{Q,i} |V_{ji}|^2 = P_j, \quad j = 1 \ldots n$$

where $V_{ji}$ denotes the $(j,i)$ entry of $\mathbf{V_H}$. Now form a new $n \times n$ matrix $\mathbf{W}$ with elements $W_{ji} = |V_{ji}|^2$ and express the eigenvalues of $\mathbf{Q}$ in a vector form as $\lambda_{\mathbf{Q}}$, then we obtain

$$\mathbf{W} \lambda_{\mathbf{Q}} = \mathbf{p}, \tag{7}$$

where $\mathbf{p} = [P_1 \ldots P_n]^T$. (Here $\mathbf{p}$ is a vector containing the diagonal values of $\mathbf{P}$.)

If the power constraint for each antenna is the same, that is $\mathbf{p} = \frac{P}{n}\mathbf{1}$, the above equation has a unique solution of $\lambda_{\mathbf{Q}} = \frac{P}{n}$. This implies $\mathbf{Q} = \frac{P}{n}\mathbf{I}_n$, which is the same as the solution with multiple access constraint. Thus forcing the beam directions to be $\mathbf{V_H}$ in this case is the same as sending independent signals from different antennas, i.e., no input optimization.

For any other $\mathbf{p}$, if $\mathbf{W}$ is full-rank then the eigenvalues can be found as $\lambda_{\mathbf{Q}} = \mathbf{W}^{-1}\mathbf{p}$. The problem, however, is that the obtained $\lambda_{\mathbf{Q}}$ may be negative, thus $\mathbf{Q}$ may be non-positive semidefinite. In other words, a solution may not exist. Illustration can be found in Figure 3 of the numerical section for a $2 \times 2$ channel, in which for the infeasible cases (equation (7) does not admit non-negative solution), the obtained transmission rate is zero.

## IV. Capacity optimization with perfect CSIT

In this section, we analyze the optimization problem of finding MIMO capacity with channel known at both the transmitter and receiver. For all stated power constraints, this capacity optimization can be cast as follows.

$$\max \quad \log \det \left( \mathbf{I}_m + \mathbf{HQH}^\dagger \right) \qquad (8)$$
$$\text{s.t.} \quad g(\mathbf{Q}, \mathbf{P}) \leq 0$$
$$\mathbf{Q} \succeq 0,$$

where $g(\mathbf{Q}, \mathbf{P}) \leq 0$ refers to a power constraint as in (3), (4) or (5), and $\mathbf{Q}$ is Hermitian.

Since all considered power constraints are linear in $\mathbf{Q}$, the above optimization is convex with any power constraint. By Slater's condition [6], because of the strictly feasible value $\mathbf{Q} = \mathbf{P} \succ 0$ which readily satisfies all power constraints, the optimal solution always exists. Thus for each power constraint, the problem admits a unique optimal solution for $\mathbf{Q}$. For the per-antenna constraint (5), this optimal $\mathbf{Q}$ is not yet known.

While the convex structure of this problem allows the optimal solution to be found numerically using available convex optimization software, such a numerical solution may be too complex for real-time system implementation and offers little analytical insights. Next, we will analyze this problem and the optimality conditions to find the optimal solution analytically. In these analyses, we assume that the channel $\mathbf{H}$ is full-rank.

### A. Capacity optimization using SDP framework

Problem (8) can be analyzed using the SDP framework. Let $\mathbf{D} = \text{diag}\{d_i\} \succeq 0$ be a diagonal matrix consisting of Lagrangian multipliers $d_i$ for the per-antenna power constraints in (5), and $\mathbf{M} \succcurlyeq 0$ be the Lagrangian multiplier for the positive semi-definite constraint. Both $\mathbf{D}$ and $\mathbf{M}$ have size $n \times n$. Then the Lagrangian for problem (8) with per-antenna power can be formed as

$$\mathcal{L}(\mathbf{Q}, \mathbf{D}, \mathbf{M}) = \log \det \left( \mathbf{I}_m + \mathbf{HQH}^\dagger \right) - \operatorname{tr}[\mathbf{D}(\mathbf{Q} - \mathbf{P})] + \operatorname{tr}(\mathbf{MQ}). \tag{9}$$

Note that we can form similar Lagrangian for the problem with sum power or multiple access constraint by replacing $\mathbf{D}$ with $\nu \mathbf{I}_n$ (scaled identity) or $\mathbf{B}$ (full matrix), respectively. These facts will be useful later when we compare solutions of different power constraints in Section VI-D. Now we will focus on per-antenna power.

Taking the first order derivative of $\mathcal{L}$ in (9) with respect to $\mathbf{Q}$ (see [7] Appendix A.7 for derivatives of a function with respect to a matrix) and equating to zero, we obtain

$$\mathbf{H}^\dagger \left( \mathbf{I}_m + \mathbf{HQH}^\dagger \right)^{-1} \mathbf{H} - \mathbf{D} + \mathbf{M} = 0.$$

Based on the KKT conditions, we then obtain a set of optimality conditions as follows.

$$\mathbf{H}^\dagger \left( \mathbf{I}_m + \mathbf{HQH}^\dagger \right)^{-1} \mathbf{H} = \mathbf{D} - \mathbf{M} \tag{10}$$

$$\mathbf{MQ} = 0$$

$$\text{diagonal } \mathbf{D} \succ 0$$

$$\text{Hermitian } \mathbf{M}, \mathbf{Q} \succcurlyeq 0.$$

Since the problem is convex, the optimal $\mathbf{Q}$ is the solution to the above set of equations. Next, we will analyze this set of equations to first deduce a condition on the rank of optimal $\mathbf{Q}$, then provide an equation for solving for $\mathbf{Q}$.

### B. Rank of the optimal input covariance

Following arguments similar to [5] (Appendix B), multiplying both sides of the first equation in (10) on the right with $\mathbf{Q}$ and applying the complementary slackness condition $\mathbf{MQ} = 0$, we obtain

$$\mathbf{DQ} = \mathbf{H}^\dagger \left( \mathbf{I}_m + \mathbf{HQH}^\dagger \right)^{-1} \mathbf{HQ}. \tag{11}$$

At optimum, we must have $\mathbf{D} \succ 0$. This is because each constraints in (5) must be met with equality, for otherwise we can always increase a diagonal value of $\mathbf{Q}$ and get a higher rate; hence the associated dual variables are strictly positive. Thus at optimum, $\mathbf{D}$ is full-rank, subsequently (11) implies that

$$\operatorname{rank}(\mathbf{Q}) \leq \operatorname{rank}(\mathbf{H}).$$

Therefore, the rank of the optimal input covariance is no more than the channel rank. In other words, the number of independent signal streams (or modes) should be no more than the rank of the channel. This result is similar to that with sum power constraint.

Since channel $\mathbf{H}$ can support at most $r = \min\{m, n\}$ independent modes (independent signal streams), the above condition implies that the rank of $\mathbf{Q}$ is at most $r$. When the rank of $\mathbf{Q}$ is less than $r$, that implies mode-dropping (similar to the same concept with sum power constraint). Since $\mathbf{QM} = 0$, $\mathbf{M}$ is a positive semidefinite matrix in the null space of $\mathbf{Q}$. The rank of $\mathbf{M}$ corresponds to the number of modes that has to be dropped for $\mathbf{Q}$ to be positive semidefinite. Suppose that the optimal solution has $k$ modes dropped ($0 \le k < \min\{m, n\}$), then

$$\text{rank}(\mathbf{M}) = k, \quad \text{rank}(\mathbf{Q}) = \min\{m, n\} - k. \tag{12}$$

The difference between the rank of $\mathbf{Q}$ and the size of $\mathbf{Q}$ should be stressed here. The size of $\mathbf{Q}$ is $n \times n$. If $n > m$ (more transmit than receive antennas), the optimal $\mathbf{Q}$ is inherently rank-deficient. In this case, even without any mode-dropping, the maximum rank of $\mathbf{Q}$ is $m < n$. Thus no mode-dropping does not always imply full-rank $\mathbf{Q}$. Only if $n \le m$ then $\mathbf{Q}$ can be full-rank without mode-dropping.

### C. Optimality conditions with per-antenna power

From the set of optimality conditions (10), we can obtain the following lemma.

**Lemma 1.** *As* $\mathbf{D}$ *is full-rank and invertible, denote* $\check{\mathbf{D}} = \mathbf{D}^{-1}$ *and define*

$$\mathbf{R}_m = \mathbf{HQH}^\dagger$$
$$\mathbf{F}_m = \mathbf{H}\check{\mathbf{D}}\mathbf{H}^\dagger, \tag{13}$$

*then the optimality conditions* (10) *imply*

$$(\mathbf{R}_m - \mathbf{F}_m + \mathbf{I}_m)\mathbf{R}_m = 0. \tag{14}$$

The proof is given in Appendix A. Note that both $\mathbf{R}_m$ and $\mathbf{F}_m$ are $m \times m$ Hermitian matrices and the achievable rate for each channel state in (2) can now be expressed as $r = \log \det (\mathbf{I}_m + \mathbf{R}_m)$, which is a sole function of $\mathbf{R}_m$.

Condition (14) provides the equation for solving for $\mathbf{Q}$. To understand the meaning of this equation better, lets denote

$$\mathbf{S}_m = \mathbf{R}_m - \mathbf{F}_m + \mathbf{I}_m. \tag{15}$$

Then we can show that (see Appendix A)

$$\mathbf{H}\check{\mathbf{D}}\mathbf{M} = \mathbf{S}_m\mathbf{H}. \tag{16}$$

Since $\check{\mathbf{D}}$ is square and full rank, (16) implies $\text{rank}(\mathbf{S}_m) = \text{rank}(\mathbf{M}) = k$. From (12) and (13), we have $\text{rank}(\mathbf{R}_m) = \text{rank}(\mathbf{Q}) = \min\{m, n\} - k$. From (14), we have $\mathbf{S}_m\mathbf{R}_m = 0$. Thus $\mathbf{S}_m$ is a matrix in the null space of $\mathbf{R}_m$, in the same way that $\mathbf{M}$ is in the null space of $\mathbf{Q}$ (but not necessarily spanning all the null space). In other words, $\mathbf{R}_m$ contains the active transmission modes, and $\mathbf{S}_m$ contains the modes that are dropped. Equation (14) essentially transforms the slackness condition from $(\mathbf{M}, \mathbf{Q})$ space to $(\mathbf{S}_m, \mathbf{R}_m)$ space.

Next, we will use (14) to solve for the optimal $\mathbf{Q}$.

## V. Optimal input covariance with per-antenna power

In this section, we establish the optimal value of $\mathbf{Q}$ as an explicit function of the dual variable $\mathbf{D}$, using Lemma 1. For the preliminary, let the singular value decomposition of the channel matrix be

$$\mathbf{H} = \mathbf{U_H} \mathbf{\Sigma_H} \mathbf{V_H}^{\dagger} \tag{17}$$

where

- $\mathbf{U_H}$ is a $m \times m$ unitary matrix containing the left singular vectors,
- $\mathbf{V_H}$ is a $n \times n$ unitary matrix containing the *right singular vectors* and
- $\mathbf{\Sigma_H}$ is a $m \times n$ diagonal matrix containing the (real) singular values in decreasing order.

It is now necessary to distinguish two cases of channel sizes: $n \leq m$ and $n > m$.

### A. Case $n \leq m$ (fewer transmit than receive antennas)

For $n \leq m$, the optimal $\mathbf{Q}$ solution can have up to all $n$ modes. In Lemma 1, however, $\mathbf{F}_m$ is not full-rank. To proceed, we need to convert this matrix to full-rank as follows.

In (17), for $n \leq m$, we can write $\mathbf{\Sigma_H} = [\mathbf{\Sigma}_n \ \mathbf{0}_{n,m-n}]^T$, where $\mathbf{\Sigma}_n$ is a $n \times n$ diagonal matrix containing the real (non-zero) singular values of $\mathbf{H}$. Now define

$$\mathbf{K} = \mathbf{V_H} \mathbf{\Sigma}_n \mathbf{V_H}^{\dagger}, \tag{18}$$

then $\mathbf{K}$ is square and full-rank. From Lemma 1, we can derive the following result.

**Lemma 2.** *Define two $n \times n$ matrices as*

$$\mathbf{R}_n = \mathbf{KQK}^{\dagger}$$

$$\mathbf{F}_n = \mathbf{K}\breve{\mathbf{D}}\mathbf{K}^{\dagger}. \tag{19}$$

*then for $n \leq m$, the optimality conditions* (10) *imply*

$$(\mathbf{R}_n - \mathbf{F}_n + \mathbf{I}_n)\mathbf{R}_n = 0. \tag{20}$$

We get an equation similar to (14), but here for $n \leq m$, $\mathbf{F}_n$ is full-rank.

*Proof:* Multiplying (14) on the left with $\mathbf{H}^\dagger$ and on the right with $\mathbf{H}$, and noting that $\mathbf{H}^\dagger\mathbf{H} = \mathbf{K}\mathbf{K}^\dagger$, we obtain

$$\mathbf{K}\left[(\mathbf{R}_n - \mathbf{F}_n + \mathbf{I}_n)\mathbf{R}_n\right]\mathbf{K}^\dagger = 0.$$

Since for $n \leq m$, $\mathbf{K}$ is square, full-rank and hence is invertible, the above equation is equivalent to (20). ∎

We now analyze equation (20). This equation can be written as $\mathbf{R}_n^2 + \mathbf{R}_n = \mathbf{F}_n\mathbf{R}_n$. This equality implies that $\mathbf{F}_n\mathbf{R}_n$ is Hermitian and has the same eigenvalue decomposition as $\mathbf{R}_n^2 + \mathbf{R}_n$, which has the same eigenvectors as those of $\mathbf{R}_n$. This is possible only if $\mathbf{R}_n$ and $\mathbf{F}_n$ share the same eigenvectors for the non-zero eigenvalues. Now since for $n \leq m$, $\mathbf{F}_n$ is full rank and Hermitian, it has a unique eigenvalue decomposition (up to any multiplicity of eigenvalues). From (19), rank($\mathbf{R}_n$) = rank($\mathbf{Q}$) = $n - k$. Equation (20) then implies that $\mathbf{R}_n$ must span $n - k$ eigenspaces of $\mathbf{F}_n$: specifically, the $n - k$ eigenvectors corresponding to the non-zero eigenvalues of $\mathbf{R}_n$ are the same as $n - k$ eigenvectors of $\mathbf{F}_n$. Define

$$\mathbf{S}_n = \mathbf{R}_n - \mathbf{F}_n + \mathbf{I}_n, \tag{21}$$

then equivalently, (20) implies that $\mathbf{S}_n$ spans the *other* $k$ eigenspaces of $\mathbf{F}_n$: the $k$ eigenvectors with non-zero eigenvalues of $\mathbf{S}_n$ are the same as the other $k$ eigenvectors of $\mathbf{F}_n$.

Intuitively, this result can be interpreted as follows. Note that we can write $\mathbf{F}_n - \mathbf{I}_n = \mathbf{R}_n - \mathbf{S}_n$. The matrix $\mathbf{F}_n - \mathbf{I}_n$ may contain some positive and some non-positive eigenvalues. Then $\mathbf{R}_n$ is the portion that contains only the positive eigenmodes, and $(-\mathbf{S}_n)$ is the portion that contains only the non-positive eigenmodes. As such, both $\mathbf{R}_n$ and $\mathbf{S}_n$ are positive semidefinite matrices and are orthogonal to each other. Here $\mathbf{R}_n$ contains the (positive) $n - k$ transmission modes, while $\mathbf{S}_n$ contains the $k$ modes that are dropped.

Based on this analysis, we can obtain the optimal value for $\mathbf{Q}$ as a function of $\mathbf{D}$ as follows.

**Theorem 1.** *For $n \leq m$, $\mathbf{K}$ as defined in* (18) *is full-rank and invertible. Denote $\check{\mathbf{K}} = \mathbf{K}^{-1}$, then for a given $\mathbf{D} \succ 0$, the optimal $\mathbf{Q}$ satisfying the optimality conditions* (10) *is given by*

$$\mathbf{Q}^\star = \check{\mathbf{D}} - \check{\mathbf{K}}\check{\mathbf{K}}^\dagger + \mathbf{Z}, \tag{22}$$

*where $\mathbf{Z} = \check{\mathbf{K}}\mathbf{S}_n\check{\mathbf{K}}^\dagger$, and $(-\mathbf{S}_n)$ is obtained as the non-positive eigenmodes of $\mathbf{K}\check{\mathbf{D}}\mathbf{K}^\dagger - \mathbf{I}_n$.*

*Proof:* From (19), we have $\mathbf{Q} = \check{\mathbf{K}}\mathbf{R}_n\check{\mathbf{K}}^\dagger$. Result (22) then follows directly from (20), (21) and the associated analysis. ∎

Theorem 1 gives the solution for $\mathbf{Q}$ in terms of the dual variable $\mathbf{D}$. Here $\mathbf{K}$ is a function of the channel as defined in (18), while $\mathbf{S}_n$ is determined from $\mathbf{F}_n$ which is a function of $\mathbf{D}$ as in (19). Note that since $\mathbf{R}_n$ contains only the positive eigenmodes of $\mathbf{F}_n - \mathbf{I}_n$, the optimal $\mathbf{Q}^\star$ as formed in (22) is always positive semidefinite. Thus the only step left is to find the optimal dual variable $\mathbf{D}$ such that $\mathbf{Q}^\star$ satisfies the power constraint of $\text{diag}(\mathbf{Q}^\star) = \mathbf{P}$.

To find the optimal $\mathbf{D}$, at this point, we need to use an iterative algorithm which we will discuss in Section VI.

### B. Case $n > m$ (more transmit than receive antennas)

For $n > m$, the optimal $\mathbf{Q}$ is inherently rank-deficient since the channel can support at most $m$ modes, which is smaller than the number of transmit antennas. For this case, we need to further decompose channel $\mathbf{H}$ as follows.

In (17), for $n > m$, we can write $\boldsymbol{\Sigma}_\mathbf{H} = [\boldsymbol{\Sigma}_m\ \mathbf{0}_{m,n-m}]$, where $\boldsymbol{\Sigma}_m$ is a $m \times m$ diagonal matrix containing the singular values of $\mathbf{H}$. Now separate the right singular vectors of $\mathbf{H}$ as

$$\mathbf{V}_\mathbf{H} = [\mathbf{V}_1\ \mathbf{V}_2], \quad \mathbf{V}_1 = \text{ first } m \text{ columns} \qquad (23)$$
$$\mathbf{V}_2 = \text{ last } n - m \text{ columns}.$$

Here $\mathbf{V}_1$ contains the basis for the row space of $\mathbf{H}$, while $\mathbf{V}_2$ contains the basis for the null space of $\mathbf{H}$. Only $\mathbf{V}_1$ is unique, but $\mathbf{V}_2$ can be any basis matrix spanning the null space of $\mathbf{H}$ (i.e. of $\mathbf{V}_1$). Note that we can also write $\mathbf{H} = \mathbf{U}_\mathbf{H}\boldsymbol{\Sigma}_m\mathbf{V}_1^\dagger$. Now lets denote the channel "inverse" $\check{\mathbf{H}}$ as

$$\check{\mathbf{H}} = \mathbf{V}_1\boldsymbol{\Sigma}_m^{-1}\mathbf{U}_\mathbf{H}^\dagger, \qquad (24)$$

then $\mathbf{H}\check{\mathbf{H}} = \mathbf{I}_m$. The use of $\check{\mathbf{H}}$ will become apparent later.

Next, from (10), multiplying both sides of the first equation on the left with $\mathbf{V}_2^\dagger$ and on the right with $\mathbf{Q}$, then applying the second equation, we get

$$\mathbf{V}_2^\dagger\mathbf{D}\mathbf{Q} = 0. \qquad (25)$$

Equation (25) places a constraint on the rank of $\mathbf{Q}$ as a direct consequence of $m < n$.

In this case, $\mathbf{F}_m$ as defined in (13) is full rank and Hermitian, hence it has unique eigenvalue decomposition (up to any multiplicity of eigenvalues). Equation (14) of Lemma 1 then implies $\mathbf{R}_m$ and $\mathbf{S}_m$ share eigenvectors with $\mathbf{F}_m$, where $\mathbf{R}_m$ spans $m-k$ eigenspaces of $\mathbf{F}_m$, and $\mathbf{S}_m$ spans the other $k$ eigenspaces of $\mathbf{F}_m$. Specifically, $\mathbf{R}_n$ contains the positive eigenmodes of $\mathbf{F}_m - \mathbf{I}_m$ while $(-\mathbf{S}_m)$ contains the non-positive eigenmodes. (The negation is just for convenience so that $\mathbf{S}_m$ is positive semidefinite.)

Now from (13) and (15), we obtain

$$\mathbf{H}(\mathbf{Q} - \check{\mathbf{D}})\mathbf{H}^\dagger = \mathbf{S}_m - \mathbf{I}_m. \qquad (26)$$

Different from (20) of Lemma 2, the above equation is under-determined for $\mathbf{Q}$. In order to find the unique optimal $\mathbf{Q}^\star$, we need to combine (26) with the rank condition (25). Based on these two equations, we can obtain the optimal $\mathbf{Q}^\star$ as follows.

**Theorem 2.** *For $n > m$, establish $\mathbf{V}_1, \mathbf{V}_2$ as in (23) and $\check{\mathbf{H}}$ as in (24). Then for a given $\mathbf{D} \succ 0$, the optimal $\mathbf{Q}$ satisfying the optimality conditions (10) is given by*

$$\mathbf{Q}^\star = \check{\mathbf{D}} - \check{\mathbf{H}}\check{\mathbf{H}}^\dagger + \mathbf{Z} - \mathbf{X}, \qquad (27)$$

*where $\mathbf{Z} = \check{\mathbf{H}}\mathbf{S}_m\check{\mathbf{H}}^\dagger$, and $(-\mathbf{S}_m)$ is obtained as the non-negative eigenmodes of $\mathbf{H}\check{\mathbf{D}}\mathbf{H}^\dagger - \mathbf{I}_m$. Here $\mathbf{X}$ is a Hermitian matrix given as*

$$\mathbf{X} = \mathbf{V}_2\mathbf{A}\mathbf{V}_2^\dagger + \mathbf{V}_1\mathbf{B}\mathbf{V}_2^\dagger + \mathbf{V}_2\mathbf{B}^\dagger\mathbf{V}_1^\dagger, \qquad (28)$$

*where $\mathbf{A}$ is a $(n-m) \times (n-m)$ Hermitian matrix and $\mathbf{B}$ is a $m \times (n-m)$ matrix given by*

$$\mathbf{B} = \mathbf{V}_1^\dagger\left(\mathbf{Z} - \check{\mathbf{H}}\check{\mathbf{H}}^\dagger\right)\mathbf{D}\mathbf{V}_2\left(\mathbf{V}_2^\dagger\mathbf{D}\mathbf{V}_2\right)^{-1},$$

$$\mathbf{A} = \left(\mathbf{I}_{n-m} - \mathbf{B}^\dagger\mathbf{V}_1^\dagger\mathbf{D}\mathbf{V}_2\right)\left(\mathbf{V}_2^\dagger\mathbf{D}\mathbf{V}_2\right)^{-1}. \qquad (29)$$

The proof is in Appendix B.

Equations (27)–(29) of Theorem 2 give the solution for $\mathbf{Q}$ in terms of the dual variable $\mathbf{D}$ in a form similar to (22) of Theorem 1. However, there are extra terms here involving $\mathbf{V}_2$ which spans the null space of the channel. In (27) and (29), $\mathbf{Z}$ is a function of $\mathbf{S}_m$ which is determined from $\mathbf{F}_m$, which in turn is a function of $\mathbf{D}$. Thus again the only remaining step is to find a diagonal $\mathbf{D} \succ 0$ such that $\mathrm{diag}(\mathbf{Q}^\star) = \mathbf{P}$.

*C. The duality gap*

To better understand the results of Theorems 1 and 2, we now analyze the duality gap. The solutions of $\mathbf{Q}^\star$ in (22) and (27) always satisfy $\mathbf{M}\mathbf{Q} = 0$ and $\mathbf{Q} \succcurlyeq 0$ as depicted in the set of optimality conditions (10). For a given dual variable $\mathbf{D}$, such a $\mathbf{Q}^\star$ is precisely the optimal primal variable to establish the Lagrange dual function from the Lagrangian (9) as

$$\mathcal{L}^\star(\mathbf{D}) = \max_{\mathbf{Q},\mathbf{M}} \mathcal{L}(\mathbf{Q}, \mathbf{D}, \mathbf{M})$$

$$= \log\det\left(\mathbf{I}_n + \mathbf{H}\mathbf{Q}^\star\mathbf{H}^\dagger\right) - \mathrm{tr}[\mathbf{D}(\mathbf{Q}^\star - \mathbf{P})].$$

The optimal dual variable $\mathbf{D}^\star$ is then the value that minimizes the above dual function. Solving for $\mathbf{D}^\star$ by directly minimizing $\mathcal{L}^\star(\mathbf{D})$, however, is not simple because the derivative of $\mathbf{Q}^\star$ with respect to $\mathbf{D}$ is complicated and may not even be derivable in closed-form. However, since problem (8) is convex and satisfies Slater's condition, it has zero duality gap at optimum. For a given $\mathbf{D}$, the duality gap is

$$\mathcal{G}(\mathbf{D}) = -\mathrm{tr}[\mathbf{D}(\mathbf{Q}^\star - \mathbf{P})]. \qquad (30)$$

Any algorithm that has this duality gap approach zero will converge to the optimal value.

## VI. Algorithm for finding the optimal $\mathbf{Q}$

In this section, we investigate the remaining question of finding a diagonal matrix $\check{\mathbf{D}} \succ 0$ such that the solutions in Theorems 1 and 2 satisfy $\operatorname{diag}(\mathbf{Q}^\star) = \mathbf{P}$. There appears to be no closed-form analytical solution for such a $\check{\mathbf{D}}$. Fortunately, equations (22) and (27) in these theorems suggest a way to compute $\mathbf{D}$ iteratively. In the follows, we design an iterative algorithm for finding the optimal $\check{\mathbf{D}}$, and hence optimal $\mathbf{Q}^\star$, for each case of $n \leq m$ and $n > m$. Then we integrate both cases in a main program for any channel size.

### A. Iterative algorithm for finding $\mathbf{Q}^\star$ when $n \leq m$

Equation (22) in Theorem 1 suggests a simple iterative algorithm for finding the optimal $\check{\mathbf{D}}$.

First we need to choose an initial point $\check{\mathbf{D}}_0$. This point can be chosen arbitrarily as long as it satisfies $\check{\mathbf{D}}_0 \succ 0$. For potentially faster convergence, we follow the mode-dropping approach similar to water filling. For the initial point, we assume that there is no mode-dropping. Denote $\check{\mathbf{G}} = \check{\mathbf{K}}\check{\mathbf{K}}^\dagger$, then based on (22), we can just simply choose diagonal matrix $\check{\mathbf{D}}_0$ as

$$\check{\mathbf{D}}_0 = \mathbf{P} + \operatorname{diag}(\check{\mathbf{G}}). \tag{31}$$

This solution always satisfies $\mathbf{D} \succ 0$ since $\check{\mathbf{G}}$ as a positive semidefinite matrix has non-negative diagonal values. At this step, we can perform a quick check to see if $\check{\mathbf{D}}_0 - \check{\mathbf{G}}$ is positive semidefinite. If it is, then this value is the optimal input covariance, i.e. $\mathbf{Q}^\star = \check{\mathbf{D}}_0 - \check{\mathbf{G}}$, and no iteration is needed.

If $(\check{\mathbf{D}}_0 - \check{\mathbf{G}})$ is non-positive semidefinite, then we adjust $\check{\mathbf{D}}$ using an iterative procedure as follows. Say at iteration $i$ ($i \geq 0$), we have obtained $\check{\mathbf{D}}_i$. Then we can form $\mathbf{F}_{n,i}$, $\mathbf{S}_{n,i}$ and $\mathbf{Q}_i$ as

$$\mathbf{F}_{n,i} = \mathbf{K}\check{\mathbf{D}}_i\mathbf{K}^\dagger$$
$$-\mathbf{S}_{n,i} = \text{ non-positive eigenmodes of } (\mathbf{F}_{n,i} - \mathbf{I}_n)$$
$$\mathbf{Z}_i = \check{\mathbf{K}}\mathbf{S}_{n,i}\check{\mathbf{K}}^\dagger$$
$$\mathbf{Q}_i = \check{\mathbf{D}}_i - \check{\mathbf{G}} + \mathbf{Z}_i. \tag{32}$$

The $\mathbf{Q}_i$ as computed in (32) is always positive semidefinite (as a consequence of Theorem 1). The term $\mathbf{Z}_i \succcurlyeq 0$ can be thought of as the adjustment at step $i$ to make $\mathbf{Q}_i \succcurlyeq 0$. But it does not guarantee that the diagonal of $\mathbf{Q}_i$ will be $\mathbf{P}$. From (32), noting that $\check{\mathbf{D}}_i$ is diagonal, we update $\check{\mathbf{D}}_{i+1}$ by the difference between the diagonal of $\mathbf{Q}_i$ and $\mathbf{P}$ as

$$\check{\mathbf{D}}_{i+1} = \check{\mathbf{D}}_i + \mathbf{P} - \operatorname{diag}(\mathbf{Q}_i). \tag{33}$$

Iteration stops when the diagonal values of $\mathbf{Q}_i$ is close to $\mathbf{P}$ within an acceptable tolerance. In implementation, we choose to stop when the duality gap (30) satisfies $|\operatorname{tr}[\mathbf{D}(\mathbf{Q}-\mathbf{P})]| < \epsilon$. Since

---

**Algorithm 1** *drop-rank-n*$(n, \check{\mathbf{D}}_0, \mathbf{K}, \check{\mathbf{K}}, \check{\mathbf{G}}, \mathbf{P}, \epsilon)$: Iterative search for $\mathbf{Q}^\star$ when $n \leq m$.

---

**Require:** $\check{\mathbf{D}}_0$ diagonal $\succ 0$, $\mathbf{P}$ diagonal $\succ 0$, $\epsilon > 0$. Also $\check{\mathbf{K}} = \mathbf{K}^{-1}$ and $\check{\mathbf{G}} = \check{\mathbf{K}}\check{\mathbf{K}}^\dagger$.

1:  Initialize $i = 0$. (iteration count)

2:  Initialize $\Delta = 1 + \epsilon$. (loop terminating variable)

3:  **while** $(\Delta > \epsilon)$ **do**

4:      Form $\mathbf{F}_i = \mathbf{K}\check{\mathbf{D}}_i\mathbf{K}^\dagger - \mathbf{I}_n$. (note that $-\mathbf{I}_n$ is included here)

5:      Perform the eigenvalue decomposition $\mathbf{F}_i = \mathbf{U}_{\mathbf{F}}\mathbf{\Lambda}_{\mathbf{F}}\mathbf{U}_{\mathbf{F}}^\dagger$. Let $k$ be the number of non-positive eigenvalues of $\mathbf{F}_i$.

6:      Form $\mathbf{S}_i = -\mathbf{U}_{\mathbf{F}}^k\mathbf{\Lambda}_{\mathbf{F}}^k\mathbf{U}_{\mathbf{F}}^{k\,\dagger}$ where

       $\mathbf{\Lambda}_{\mathbf{F}}^k$ is the $k \times k$ diagonal matrix of all $k$ non-positive eigenvalues of $\mathbf{F}_i$,

       $\mathbf{U}_{\mathbf{F}}^k$ consists of the corresponding $k$ eigenvectors.

7:      Form $\mathbf{Z}_i = \check{\mathbf{K}}\mathbf{S}_i\check{\mathbf{K}}^\dagger$.

8:      Form $\mathbf{Q}_i = \check{\mathbf{D}}_i - \check{\mathbf{G}} + \mathbf{Z}_i$.

9:      Form $\check{\mathbf{D}}_{i+1} = \check{\mathbf{D}}_i + \mathbf{P} - \text{diag}(\mathbf{Q}_i)$.

10:    Compute $\Delta = |\text{tr}[\mathbf{D}_i(\mathbf{Q}_i - \mathbf{P})]|$.

11:    $i \leftarrow i + 1$.

12: **end while**

13: **return** $\check{\mathbf{D}}_i$ and $\mathbf{Q}_i$.

---

problem (8) is convex and satisfies Slater's condition, this stopping criterion always guarantees the optimal solution. The iterative procedure for finding $\mathbf{Q}^\star$ when $n \leq m$ is summarized in Algorithm 1, *drop-rank-n*$(\cdot)$.

### B. Iterative algorithm for finding $\mathbf{Q}^\star$ when $n > m$

For $n > m$, we rely on equation (27) in Theorem 2 to design a similar algorithm. Let $\check{\mathbf{G}} = \check{\mathbf{H}}\check{\mathbf{H}}^\dagger$, (note that we reuse the symbol $\check{\mathbf{G}}$ here but there should no confusion based on the channel size). Again we can start with the initial value $\check{\mathbf{D}}_0$ in (31) or with any arbitrary diagonal $\check{\mathbf{D}}_0 \succ 0$. In this case, however, iteration is always necessary (except in the unlikely event that the algorithm starts with the *optimal* $\mathbf{D}$).

---

**Algorithm 2** *drop-rank-m*$(m, \check{\mathbf{D}}_0, \mathbf{H}, \check{\mathbf{H}}, \check{\mathbf{G}}, \mathbf{V}_1, \mathbf{V}_2, \mathbf{P}, \epsilon)$: Iterative search for $\mathbf{Q}^\star$ when $n > m$.

**Require:** $\check{\mathbf{D}}_0$ diagonal $\succ 0$, $\mathbf{P}$ diagonal $\succ 0$, $\epsilon > 0$. Also $\mathbf{V}_1$ and $\mathbf{V}_2$ are related to $\mathbf{H}$ as in (23), $\check{\mathbf{H}}$ is defined in (24) and $\check{\mathbf{G}} = \check{\mathbf{H}}\check{\mathbf{H}}^\dagger$.

1: Initialize $i = 0$. (iteration count)

2: Initialize $\Delta = 1 + \epsilon$. (loop terminating variable)

3: **while** $(\Delta > \epsilon)$ **do**

4:     Form $\mathbf{F}_i = \mathbf{H}\check{\mathbf{D}}_i\mathbf{H}^\dagger - \mathbf{I}_m$. (note that $-\mathbf{I}_m$ is included here)

5:     Perform the eigenvalue decomposition $\mathbf{F}_i = \mathbf{U_F}\mathbf{\Lambda_F}\mathbf{U_F}^\dagger$.

6:     Form $\mathbf{S}_i = -\mathbf{U_F}^k\mathbf{\Lambda_F}^k\mathbf{U_F}^{k\,\dagger}$ that contains all $k$ non-positive eigenmodes of $\mathbf{F}_i$.

7:     Form $\mathbf{Z}_i = \check{\mathbf{H}}\mathbf{S}_i\check{\mathbf{H}}^\dagger$.

8:     Form $\mathbf{D}_i = \text{diag}\{(\check{D}_{i,jj})^{-1}\}$, $j = 1 \ldots n$.

9:     Form $\mathbf{B}_i = \mathbf{V}_1^\dagger\left(\mathbf{Z}_i - \check{\mathbf{G}}\right)\mathbf{D}_i\mathbf{V}_2\left(\mathbf{V}_2^\dagger\mathbf{D}_i\mathbf{V}_2\right)^{-1}$.

10:     Form $\mathbf{A}_i = \left(\mathbf{I}_{n-m} - \mathbf{B}_i^\dagger\mathbf{V}_1^\dagger\mathbf{D}_i\mathbf{V}_2\right)\left(\mathbf{V}_2^\dagger\mathbf{D}_i\mathbf{V}_2\right)^{-1}$.

11:     Form $\mathbf{X}_i = \mathbf{V}_2\mathbf{A}_i\mathbf{V}_2^\dagger + \mathbf{V}_1\mathbf{B}_i\mathbf{V}_2^\dagger + \mathbf{V}_2\mathbf{B}_i^\dagger\mathbf{V}_1^\dagger$.

12:     Form $\mathbf{Q}_i = \check{\mathbf{D}}_i - \check{\mathbf{G}} + \mathbf{Z}_i - \mathbf{X}_i$.

13:     Form $\check{\mathbf{D}}_{i+1} = \check{\mathbf{D}}_i + \mathbf{P} - \text{diag}(\mathbf{Q}_i)$.

14:     Compute $\Delta = |\text{tr}[\mathbf{D}_i(\mathbf{Q}_i - \mathbf{P})]|$.

15:     $i \leftarrow i + 1$.

16: **end while**

17: **return** $\check{\mathbf{D}}_i$ and $\mathbf{Q}_i$.

---

At step $i$ $(i \geq 0)$, having obtained $\check{\mathbf{D}}_i$, we compute

$$\mathbf{F}_{m,i} = \mathbf{H}\check{\mathbf{D}}_i\mathbf{H}^\dagger$$

$$-\mathbf{S}_{m,i} = \text{ non-positive eigenmodes of } (\mathbf{F}_{m,i} - \mathbf{I}_m)$$

$$\mathbf{Z}_i = \check{\mathbf{H}}\mathbf{S}_{m,i}\check{\mathbf{H}}^\dagger$$

$$\mathbf{X}_i = \mathbf{V}_2\mathbf{A}_i\mathbf{V}_2^\dagger + \mathbf{V}_1\mathbf{B}_i\mathbf{V}_2^\dagger + \mathbf{V}_2\mathbf{B}_i^\dagger\mathbf{V}_1^\dagger$$

$$\mathbf{Q}_i = \check{\mathbf{D}}_i - \check{\mathbf{G}} + \mathbf{Z}_i - \mathbf{X}_i, \tag{34}$$

where $\mathbf{A}_i$ and $\mathbf{B}_i$ are computed from $\mathbf{D}_i$ and $\mathbf{Z}_i$ as in (29). The $\mathbf{Q}_i$ as computed in (34) is again always positive semidefinite and has rank at most $m$. Then we update $\check{\mathbf{D}}_{i+1}$ in the same way as in (33), and stop when the duality gap $|\text{tr}[\mathbf{D}(\mathbf{Q} - \mathbf{P})]|$ is sufficiently small. This procedure is summarized in Algorithm 2, *drop-rank-m*$(\cdot)$.

**Algorithm 3** *opt-cov*($\mathbf{H}, \{P_i\}, \epsilon$): The main program to find $\mathbf{Q}^\star$ for a given channel $\mathbf{H}$.

**Require:** Channel $\mathbf{H}$ full-rank, per-antenna power $P_i$ real and $P_i \geq 0$, $i = 1 \ldots n$, precision $\epsilon > 0$.

1: Let $(m, n) = \text{size}(\mathbf{H})$,

   $m$ = number of rows or receive antennas,

   $n$ = number of columns or transmit antennas.

2: Perform the singular value decomposition $\mathbf{H} = \mathbf{U_H} \mathbf{\Sigma_H} \mathbf{V_H^\dagger}$, where

   $\mathbf{U_H}$: $m \times m$ unitary matrix,

   $\mathbf{V_H}$: $n \times n$ unitary matrix,

   $\mathbf{\Sigma_H}$: $m \times n$ diagonal matrix with diagonal entries as real, non-zero singular values $\sigma_{\mathbf{H},i}$ in decreasing order.

3: **if** $n > m$ **then**

4:   Form $\mathbf{V}_2$ from the last $n - m$ columns of $\mathbf{V_H}$.

5:   Form $\mathbf{V}_1$ from the first $m$ columns of $\mathbf{V_H}$.

6:   Form $\mathbf{\check{\Sigma}}_m = \text{diag}\{\sigma_{\mathbf{H},i}^{-1}\}$, $m \times m$ diagonal matrix.

7:   Form $\mathbf{\check{H}} = \mathbf{V}_1 \mathbf{\check{\Sigma}}_m \mathbf{U_H}$.

8:   Form $\mathbf{\check{G}} = \mathbf{V}_1 \mathbf{\check{\Sigma}}_m^2 \mathbf{V}_1^\dagger$.

9: **else**

10:   Form $\mathbf{\Sigma}_n = \text{diag}\{\sigma_{\mathbf{H},i}\}$, $n \times n$ diagonal matrix.

11:   Form $\mathbf{K} = \mathbf{V_H} \mathbf{\Sigma}_n \mathbf{V_H^\dagger}$.

12:   Form $\mathbf{\check{\Sigma}}_n = \text{diag}\{\sigma_{\mathbf{H},i}^{-1}\}$, $n \times n$ diagonal matrix.

13:   Form $\mathbf{\check{K}} = \mathbf{V_H} \mathbf{\check{\Sigma}}_n \mathbf{V_H^\dagger}$.

14:   Form $\mathbf{\check{G}} = \mathbf{V_H} \mathbf{\check{\Sigma}}_n^2 \mathbf{V_H^\dagger}$.

15: **end if**

16: Form $\mathbf{P} = \text{diag}\{P_i\}$, $n \times n$ diagonal matrix.

17: Form $\mathbf{\check{D}}_0 = \mathbf{P} + \text{diag}(\mathbf{\check{G}})$.

18: **if** $n \leq m$ **then**

19:   $(\mathbf{\check{D}}, \mathbf{Q}) = $ *drop-rank-n*$(n, \mathbf{\check{D}}_0, \mathbf{K}, \mathbf{\check{K}}, \mathbf{\check{G}}, \mathbf{P}, \epsilon)$

20: **else**

21:   $(\mathbf{\check{D}}, \mathbf{Q}) = $ *drop-rank-m*$(m, \mathbf{\check{D}}_0, \mathbf{H}, \mathbf{\check{H}}, \mathbf{\check{G}}, \mathbf{V}_1, \mathbf{V}_2, \mathbf{P}, \epsilon)$

22: **end if**

23: **return** $\mathbf{Q}$.

The integrated main program for finding the optimal $\mathbf{Q}^\star$ for all channel sizes is summarized in Algorithm 3, *opt-cov*($\cdot$).

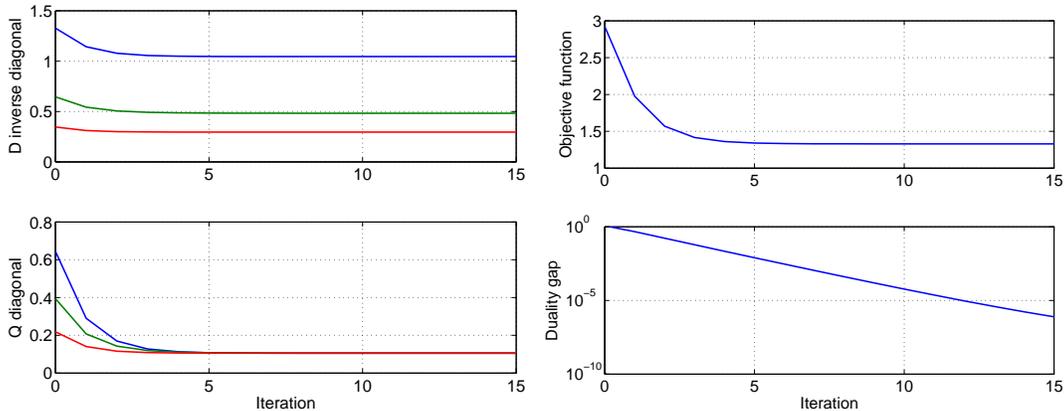

Fig. 1. Typical convergence for a $3 \times 3$ channel at SNR = -5dB and $\epsilon = 10^{-6}$.

### C. Convergence analysis

*1) For $n \leq m$:* The following Corollary shows convergence of Algorithm 1.

**Corollary 1.** *In Algorithm 1, $\check{\mathbf{D}}_i$ is decreasing in $i$. This algorithm always converges to the optimal point.*

*Proof:* Consider the iterative steps (32) and (33) with initial condition (31). We will show that $\check{\mathbf{D}}_i$ is decreasing in $i$ by induction as follows.

First, with $i = 0$, from (31) and (32) we have $\mathbf{Q}_0 = \mathbf{P} + \mathbf{Z}_0 + \text{diag}(\check{\mathbf{G}}) - \check{\mathbf{G}}$. Thus from (33),

$$\check{\mathbf{D}}_1 = \check{\mathbf{D}}_0 - \text{diag}(\mathbf{Z}_0).$$

Since $\mathbf{Z}_0 \succcurlyeq 0$, then $\text{diag}(\mathbf{Z}_0) \succcurlyeq 0$, and thus $\check{\mathbf{D}}_1 \preccurlyeq \check{\mathbf{D}}_0$.

Now assume that $\check{\mathbf{D}}_{j+1} \preccurlyeq \check{\mathbf{D}}_j$ for some $j \geq 0$. From (32), we then have $\mathbf{F}_{n,j+1} \preccurlyeq \mathbf{F}_{n,j}$. Since $\mathbf{F}_n - \mathbf{I}_n = \mathbf{R}_n - \mathbf{S}_n$, this implies $\mathbf{S}_{n,j} - \mathbf{S}_{n,j+1} \preccurlyeq \mathbf{R}_{n,j} - \mathbf{R}_{n,j+1}$. Multiplying both sides on the left with $\mathbf{S}_{n,j}$ and noting that $\mathbf{S}_{n,j}\mathbf{R}_{n,j} = 0$, we obtain $\mathbf{S}_{n,j}\left(\mathbf{S}_{n,j} - \mathbf{S}_{n,j+1}\right) \preccurlyeq -\mathbf{S}_{n,j}\mathbf{R}_{n,j+1} \preccurlyeq 0$. Since $\mathbf{S}_{n,j} \succcurlyeq 0$, this implies $\mathbf{S}_{n,j} \preccurlyeq \mathbf{S}_{n,j+1}$. Now multiplying both sides on the left with $\mathbf{K}$ and on the right with $\mathbf{K}^\dagger$, we then have $\mathbf{Z}_j \preccurlyeq \mathbf{Z}_{j+1}$.

From (33) and (32), we can write

$$\check{\mathbf{D}}_{j+1} = \mathbf{P} + \text{diag}(\check{\mathbf{G}}) - \mathbf{Z}_j.$$

Thus with $\mathbf{Z}_j \preccurlyeq \mathbf{Z}_{j+1}$, then $\check{\mathbf{D}}_{j+2} \preccurlyeq \check{\mathbf{D}}_{j+1}$. This completes the induction.

The sequence $\check{\mathbf{D}}_i$ is decreasing in $i$ and is lower bounded as $\check{\mathbf{D}}_i \succ 0$. Thus $\check{\mathbf{D}}_i$ must converge. Based on (33), then $\mathbf{Q}_i$ must also converge to an optimal value with $\text{diag}(\mathbf{Q}) = \mathbf{P}$. In other words, the algorithm always converges to the correct optimal value. ∎

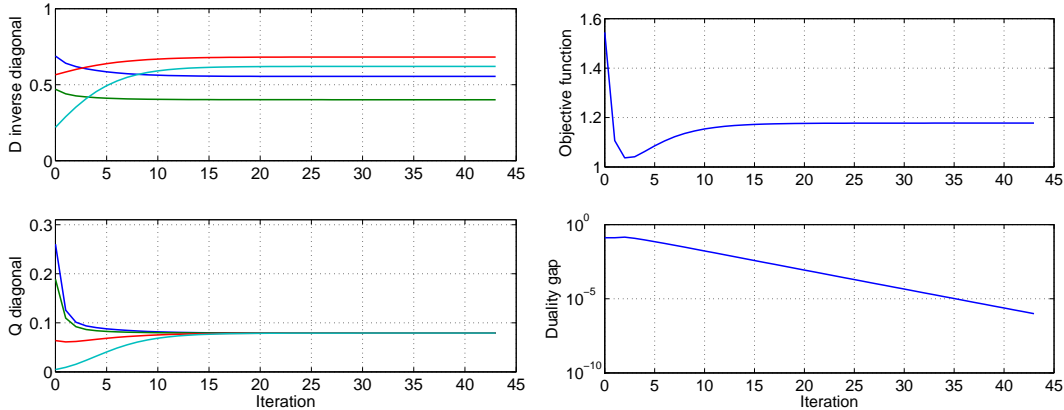

Fig. 2. Typical convergence for a $2 \times 4$ channel at SNR = -5dB, $\epsilon = 10^{-6}$ and random starting point $\check{\mathbf{D}}_0$.

Note that convergence holds for any starting point, not just the value of $\check{\mathbf{D}}_0$ in (31). Figure 1 shows a typical convergence behavior of $\check{\mathbf{D}}_i$ and $\mathrm{diag}(\mathbf{Q}_i)$ as well as the objective function value and the duality gap for $n = m = 3$ (the channel is generated randomly according to the circularly complex Gaussian distribution). The convergence rate appears to be exponentially fast.

*2) For $n > m$:* Now consider Algorithm 2 with iterative steps (34) and (33) and initial condition (31). The algorithm in this case is more complex. Here $\check{\mathbf{D}}_i$ is not always decreasing in $i$, but some of its diagonal elements are decreasing and others increasing. However, we observe that $\check{\mathbf{D}}_i \pi$ is decreasing in $i$ for some diagonal matrix $\pi$ with only 1 and $-1$ on the diagonal such that for the initial step, $\mathrm{diag}(\mathbf{Q}_0)\pi \succcurlyeq \mathbf{P}\pi$ (or after a number of iterations $K$, $\mathrm{diag}(\mathbf{Q}_K)\pi \succcurlyeq \mathbf{P}\pi$). That is $\check{\mathbf{D}}_i \pi \succcurlyeq \check{\mathbf{D}}_{i+1}\pi$. The detailed analysis is given in Appendix C.

Given $\check{\mathbf{D}}_i \pi$ is decreasing in $i$, then $\check{\mathbf{D}}_i$ converges, which implies from (33) that $\mathrm{diag}(\mathbf{Q}_i)$ converges to $\mathbf{P}$. The convergence also holds for any starting point $\check{\mathbf{D}}_0$.

Figure 2 shows a typical convergence behavior for a random channel with $n = 4$ and $m = 2$ at low SNR ($-5$dB). Again the convergence rate appears to be exponentially fast. Numerical simulations show that higher SNR generally leads to faster convergence.

### D. Connections with water-filling and multiple-access

As noted in Section IV-A, all 3 power constraints can be analyzed in the SDP framework (8). The only difference is in the dual variable associated with each power constraint in the Lagrangian (9). This dual variable for the sum, per-antenna and multiple-access constraint respectively is equal to $\nu \mathbf{I}_n$ (scaled identity), $\mathbf{D}$ (square diagonal) and $\mathbf{B}$ (square full matrix). This change of dual variable, however, makes it possible to find closed-form solutions for the sum and multiple access constraints. Next we will discuss these solutions and the implication on input eigenbeams.

*1) Optimal solutions for sum and multiple access constraints:* For the sum power constraint, all results of Theorems 1 and 2 apply by replacing $\mathbf{D}$ with $\nu \mathbf{I}_n$. This change, however, makes it possible to find closed-form solutions for $\nu$ even with mode-dropping. In particular, with sum power constraint, the result of Theorem 1 becomes

$$\mathbf{Q} = \nu^{-1}\mathbf{I}_n - \check{\mathbf{K}}\check{\mathbf{K}}^\dagger + \check{\mathbf{K}}\mathbf{S}_n\check{\mathbf{K}}^\dagger, \tag{35}$$

where $(-\mathbf{S}_n)$ now contains the non-positive eigenmodes of $\nu^{-1}\mathbf{K}\mathbf{K}^\dagger - \mathbf{I}_n$. Equation (35) then implies that $\mathbf{Q}$ has the same eigenvectors as those of $\mathbf{K}$, which are $\mathbf{V_H}$ (the right singular vectors of $\mathbf{H}$). Similar eigenvector analysis holds for Theorem 2. These eigenvectors then lead directly to the water-filling solution (6) for the eigenvalues of $\mathbf{Q}$, where $\mu = 1/\nu$. As such, a closed-form solution for the optimal dual variable $\nu^\star$ can be obtained as

$$\nu^\star = \left( P + \sum_{i=1}^{K} \frac{1}{\lambda_{\mathbf{H},i}} \right)^{-1}$$

where $K \le \min(m, n)$ is the number of active modes (assuming $\lambda_{\mathbf{H},i}$ are in decreasing order).

Similarly, for the multiple access constraint, results of Theorems 1 and 2 also hold with $\mathbf{D}$ being replaced by a full $n \times n$ positive semidefinite matrix $\mathbf{B}$. In this case, the optimal covariance is $\mathbf{Q} = \mathbf{P}$ and there is no mode dropping ($\mathbf{S}_n = 0$ and $\mathbf{S}_m = 0$). Thus for $n \le m$, we can easily identify the optimal dual variable as $\mathbf{B}^\star = (\mathbf{P} + \check{\mathbf{G}})^{-1}$. For $n > m$, closed-form expression for $\mathbf{B}^\star$ can also be derived from (26)–(29).

*2) Comments on input eigenbeams:* In both cases of sum and multiple access constraint, the eigenvectors of the optimal $\mathbf{Q}$ are independent of the dual variable. For the per-antenna constraint, however, eigenvectors of $\mathbf{Q}$ depend on the dual $\mathbf{D}$ as shown in Theorems 1 and 2. As a consequence, the optimal beamforming directions with per-antenna constraint are not the right singular vectors of the channel (as discussed in Section III). For MISO channels, the rank-one optimal $\mathbf{Q}$ is shown to have its eigenvector matched to the phase but not the amplitude of the channel vector [5]. For MIMO channels, the relation is more complicated as the eigenvectors of $\mathbf{Q}$ depend on multiple factors: the channel $\mathbf{H}$, the per-antenna constraint $\mathbf{P}$ and also the SNR. It does not appear feasible to find the optimal eigenbeams and power allocation separately. Instead, the proposed algorithms establish the optimal eigenbeams and power allocation together in the resulted optimal $\mathbf{Q}$.

## VII. FADING MIMO CAPACITIES WITH NO CSIT

In this section, we consider the case of fading channel with no CSIT. The transmitter does not know the channel realization $\mathbf{H}$ but only knows its distribution as a circularly complex

Gaussian matrix with zero mean and covariance $\mathbf{I}_{n \times m}$. In this case, we need to consider the ergodic capacity. For all power constraints, capacity optimization can now be cast as follows.

$$\max \quad E_{\mathbf{H}} \left[ \log \det \left( \mathbf{I}_m + \mathbf{HQH}^\dagger \right) \right] \tag{36}$$
$$\text{s.t.} \quad g(\mathbf{Q}, \mathbf{P}) \leq 0, \quad \mathbf{Q} \succcurlyeq 0,$$

where $g(\mathbf{Q}, \mathbf{P}) \leq 0$ refers to a specific power constraint as in (3), (4) or (5). Different from (8), the optimal $\mathbf{Q}$ is no longer a function of $\mathbf{H}$ but only of its distribution.

With sum power, the capacity of MIMO fading channel was established in the seminal paper by Telatar [2]. The optimal covariance of the Gaussian transmit signal is $\mathbf{Q} = \frac{P}{n}\mathbf{I}$, implying that each antenna sends independent signal with equal power. With multiple-access constraint, the transmit covariance again has to be $\mathbf{Q} = \mathbf{P}$.

We have established the ergodic capacity with per-antenna power for the MISO channel earlier in [5] and showed that it is the same as capacity under multiple-access constraint with optimal $\mathbf{Q} = \mathbf{P}$. The same conclusion, in fact, holds for MIMO channels. For the sake of completeness, we provide the analysis with a shorter proof for MIMO channels here. We use an approach similar to [2] but only apply the following set of diagonal matrices instead of all unitary matrices. Define $\mathbf{\Pi}^{(j)}$ as an $n \times n$ diagonal matrix with all $1$ on the diagonal except a $(-1)$ at location $j$. Then $\mathbf{H\Pi}^{(j)}$ has the same distribution as $\mathbf{H}$ and hence

$$E_{\mathbf{H}} \left[ \log \det \left( 1 + \mathbf{HQH}^\dagger \right) \right] = E_{\mathbf{H}} \left[ \log \det \left( 1 + \mathbf{H\Pi}^{(j)} \mathbf{Q} \mathbf{\Pi}^{(j)T} \mathbf{H}^\dagger \right) \right].$$

Here the set of $\mathbf{\Pi}^{(j)}$ matrices preserves the diagonal values of $\mathbf{\Pi}^{(j)} \mathbf{Q} \mathbf{\Pi}^{(j)T}$. We can then apply the following inequality based on the concavity of $\log \det$ function:

$$r = \frac{1}{n} \sum_{j=1}^{n} E_{\mathbf{H}} \left[ \log \det \left( 1 + \mathbf{H\Pi}^{(j)} \mathbf{Q} \mathbf{\Pi}^{(j)T} \mathbf{H}^\dagger \right) \right]$$
$$\leq E_{\mathbf{H}} \left[ \log \det \left( 1 + \mathbf{H} \left( \frac{1}{n} \sum_{j=1}^{n} \mathbf{\Pi}^{(j)} \mathbf{Q} \mathbf{\Pi}^{(j)T} \right) \mathbf{H}^\dagger \right) \right]$$
$$= E_{\mathbf{H}} \left[ \log \det \left( 1 + \mathbf{HPH}^\dagger \right) \right].$$

Hence the capacity is achieved by setting $\mathbf{Q} = \mathbf{P}$. Since the set of diagonal matrices $\mathbf{\Pi}^{(j)}$ used here is a subset of all unitary matrices used in [2], it follows immediately that ergodic capacity with per-antenna power is smaller than or equal to ergodic capacity with sum power. If and only if the per-antenna constraint is $\mathbf{P} = \frac{P}{n}\mathbf{I}_n$, then the two capacities are equal.

## VIII. NUMERICAL RESULTS AND ANALYSIS

In this section, we provide numerical examples to illustrate MIMO capacity with per-antenna power constraint. We will only show results for perfect CSIT, in which per-antenna power has significant impact.

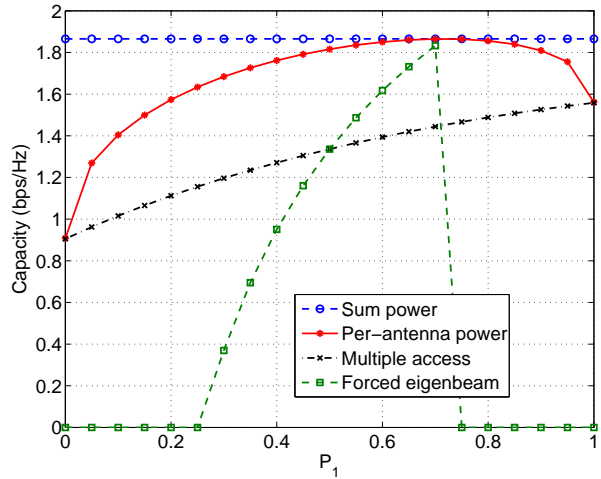

Fig. 3. Capacities of a $2 \times 2$ channel with perfect CSIT at SNR = 0dB with $\mathbf{P} = \text{diag}\{P_1, \ 1 - P_1\}$.

### A. Impact of per-antenna power for a given channel

First, to see the impact of separate power constraint on each antenna, we vary the constraint $\mathbf{P} = \text{diag}\{P_1, \ P_2\}$ for a $2 \times 2$ channel such that the total power $P_1 + P_2 = 1$ (the SNR = 0dB). Figure 3 shows a typical capacity plot for a channel randomly generated according to the circularly complex Gaussian distribution (as in Rayleigh fading). The specific channel for this figure is

$$\mathbf{H} = \left[ \begin{array}{cc} 0.0541 - 0.4066i & -0.4339 + 0.0033i \\ -1.3200 - 0.1872i & 0.8269 - 0.0279i \end{array} \right].$$

The figure shows that input power constraint can affect the capacity significantly. For example, even at equal constraints $P_1 = P_2 = 0.5$, the capacity with per antenna power is significantly higher than with multiple access constraint, but is still lower than with sum power. Sum power constraint, however, does not result in equal power at each antenna at this point.

Also plotted is the achievable rate for per-antenna constraint using forced eigenbeams as the channel right singular vectors as in (7). The 3 lines for capacity with sum power, capacity with per-antenna power and rate achievable with forced eigenbeams meet at a single point when the per-antenna constraint is such that it coincides with the optimal power per antenna under sum constraint. Other than this point, the rate with forced eigenbeams is always smaller than capacity with per-antenna power and goes to zero at certain $\mathbf{P}$ because no solutions exist.

### B. Average impact: equal or unequal constraint?

Next, to see the average impact of each power constraint, we plot the ergodic capacity for Rayleigh fading channels with perfect CSIT. Figure 4 shows the ergodic capacity of a $4 \times 2$ channel, obtained by averaging over 1000 random channel realizations at each SNR. In this figure, the per-antenna power has equal constraint at each antenna, i.e. $\mathbf{P} = \frac{P}{n}\mathbf{I}_n$. We can observe that the capacity with per-antenna power is then almost as high as that with sum power. However, the rate with forced eigenbeams is exactly the same as capacity with multiple access constraint (no input optimization), which agrees with the analysis in Section III.

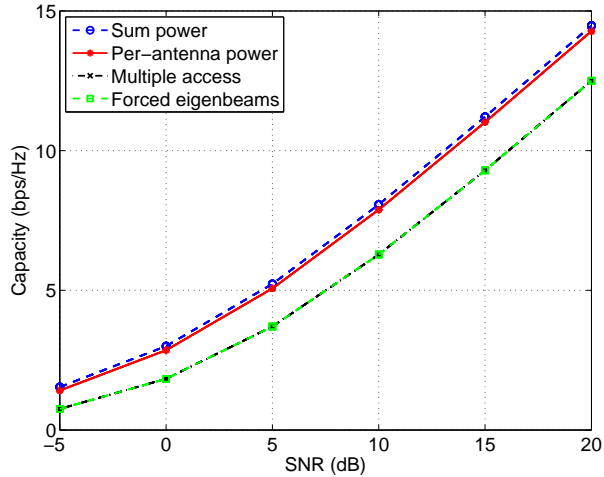

Fig. 4. Ergodic capacities of a $4 \times 2$ Rayleigh fading channel with perfect CSIT and $\mathbf{P} = \frac{P}{n}\mathbf{I}_n$. Only for this equal constraint $\mathbf{P}$ that the rate with forced eigenbeams is as large as the capacity with multiple access constraint (i.e. no input optimization).

If we make the power constraint at each antenna unequal, the difference in capacity with per-antenna power and with sum power is more pronounced, as seen in Figure 5 for a $3 \times 3$ channel with power at antenna $k$ proportional to $k^2$. The more skew the constraint, the more different the ergodic capacity becomes under per-antenna and sum power constraint.

Figure 5 also shows the rate with forced eigenbeams at almost zero for all SNR. Even though this plot emphasizes the impact by skewing the per-antenna constraint significantly, simulations show that even if the constraint is only slightly different from $\frac{P}{n}\mathbf{I}_n$, rate with forced eigenbeams falls strictly below rate with no input optimization. This relation readily follows from (7), since for $\mathbf{P} \neq \frac{P}{n}\mathbf{I}_n$, the probability that equation (7) has no non-negative solution is strictly non-zero. As the constraint becomes more skew, the average rate with forced eigenbeams approaches zero.

### C. Impact of channel size

Simulations also show that per-antenna power has higher impact on ergodic capacity when $n > m$. If the power at each antenna is constrained to be the same ($\mathbf{P} = \frac{P}{n}\mathbf{I}_n$), then for $n \leq m$, the ergodic capacities under the 3 power constraints differ only slightly at low SNR and approach each other as the SNR increases. For $n > m$, however, the ergodic capacity with per-antenna power remains significantly higher than that with multiple access and is relatively close to the capacity with sum power for all SNR.

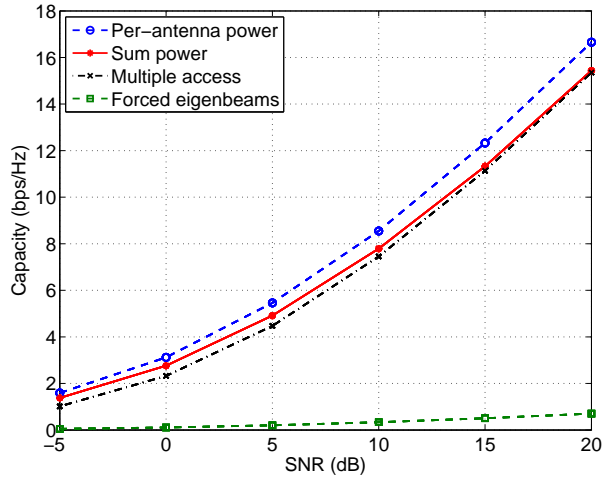

Fig. 5. Ergodic capacities of a $3 \times 3$ Rayleigh fading channel with per-antenna constraint at antenna $k$ proportional to $k^2$. The capacity with per-antenna power approaches capacity with multiple access at high SNR only for $n \leq m$, but remains significantly higher at all SNR for $n > m$. The rate with forced eigenbeam is almost zero for all channel sizes.

On the other hand, if the power at each antenna is constrained differently from each other, then capacity with per-antenna power can be significantly different from capacity with sum power for all channel sizes. As the SNR increases, capacity with per-antenna power approaches capacity with multiple access constraint for $n \leq m$, but remains higher for $n > m$.

Last, we have also verified that for the case of single receive antenna ($m = 1$), the proposed algorithm converges to the same result as the closed-form solution in [5].

## IX. CONCLUSION

We have established the MIMO capacity under per-antenna power constraint with perfect transmitter channel state information. The optimal input covariance matrix is no longer diagonalizable by the channel right singular vectors as with sum power constraint. We solve in closed-form the optimal covariance matrix as a function of the dual variable. We then design an efficient algorithm to find this dual variable and hence the optimal input signaling. Simulation results show that per-antenna constraint can impact the capacity significantly. The impact is more pronounced in channels with more transmit than receive antennas. In all cases, to maintain realistic per-antenna power requirements, solutions using channel eigenbeams produce rate equal or worse than that using independent signaling (no input optimization). Optimal signaling can achieve a capacity close to the idealistic capacity with sum constraint and significantly higher than capacity with independent signaling.

APPENDIX

### A. Proof of Lemma 1

In this proof, we make repeated use of the following identity expansion:

$$\mathbf{I}_m = \left(\mathbf{I}_m + \mathbf{HQH}^\dagger\right)\left(\mathbf{I}_m + \mathbf{HQH}^\dagger\right)^{-1}. \tag{37}$$

Note that the order of the two factors in this expansion is interchangeable.

To prove (14), multiplying the first equation in (10) on the right with $\mathbf{QH}^\dagger$ and on the left with $\mathbf{H\check{D}}$, and noting that $\mathbf{MQ} = 0$, we get

$$\mathbf{H\check{D}H}^\dagger \left(\mathbf{I}_m + \mathbf{HQH}^\dagger\right)^{-1} \mathbf{HQH}^\dagger = \mathbf{HQH}^\dagger.$$

Now subtracting both sides by $\mathbf{H\check{D}H}^\dagger$, then applying identity expansion (37), this equation simplifies to

$$-\mathbf{H\check{D}H}^\dagger \left(\mathbf{I}_m + \mathbf{HQH}^\dagger\right)^{-1} = \mathbf{HQH}^\dagger - \mathbf{H\check{D}H}^\dagger.$$

Next adding both sides with $\mathbf{I}_m$ and again using identity expansion (37), we get

$$\left(\mathbf{I}_m + \mathbf{HQH}^\dagger - \mathbf{H\check{D}H}^\dagger\right)\left(\mathbf{I}_m + \mathbf{HQH}^\dagger\right)^{-1} = \mathbf{I}_m + \mathbf{HQH}^\dagger - \mathbf{H\check{D}H}^\dagger.$$

Denote $\mathbf{S}_m = \mathbf{I}_m + \mathbf{HQH}^\dagger - \mathbf{H\check{D}H}^\dagger$ similar to (15), the above equation becomes

$$\mathbf{S}_m \left(\mathbf{I}_m - \left(\mathbf{I}_m + \mathbf{HQH}^\dagger\right)^{-1}\right) = 0.$$

Then applying identity expansion (37) once more, we obtain

$$\mathbf{S}_m \left(\mathbf{HQH}^\dagger\right)\left(\mathbf{I}_m + \mathbf{HQH}^\dagger\right)^{-1} = 0.$$

Since $\left(\mathbf{I}_m + \mathbf{HQH}^\dagger\right)^{-1}$ is a full-rank square matrix, the above equation is equivalent to (14).

To show (16), from the first equation in (10), subtracting $\mathbf{H}^\dagger\mathbf{H}$ from both sides, we have

$$\mathbf{H}^\dagger \left[\left(\mathbf{I}_m + \mathbf{HQH}^\dagger\right)^{-1} - \mathbf{I}_m\right] \mathbf{H} = \mathbf{D} - \mathbf{M} - \mathbf{H}^\dagger\mathbf{H}.$$

Using identity expansion (37), we obtain

$$\mathbf{H}^\dagger \left(\mathbf{I}_m + \mathbf{HQH}^\dagger\right)^{-1} \mathbf{HQH}^\dagger\mathbf{H} = \mathbf{H}^\dagger\mathbf{H} - \mathbf{D} + \mathbf{M}.$$

Now replacing a part of the left expression by the first equation in (10), we have

$$(\mathbf{D} - \mathbf{M})\mathbf{QH}^\dagger\mathbf{H} = \mathbf{H}^\dagger\mathbf{H} - \mathbf{D} + \mathbf{M}.$$

But $\mathbf{MQ} = 0$, hence we have

$$\mathbf{M} = \mathbf{DQH}^\dagger\mathbf{H} + \mathbf{D} - \mathbf{H}^\dagger\mathbf{H}.$$

Multiplying both sides on the left by $\mathbf{H\check{D}}$, we obtain (16).

## B. Proof of Theorem 2

From equation (26), multiplying on the left with $\check{\mathbf{H}}$ and on the right with $\check{\mathbf{H}}^{\dagger}$, we get

$$\mathbf{V}_1\mathbf{V}_1^{\dagger}(\mathbf{Q}-\check{\mathbf{D}})\mathbf{V}_1\mathbf{V}_1^{\dagger} = \check{\mathbf{H}}\mathbf{S}_m\check{\mathbf{H}}^{\dagger} - \check{\mathbf{H}}\check{\mathbf{H}}^{\dagger}$$

Denote $\mathbf{Z} = \check{\mathbf{H}}\mathbf{S}_m\check{\mathbf{H}}^{\dagger}$ and $\check{\mathbf{G}} = \check{\mathbf{H}}\check{\mathbf{H}}^{\dagger}$. Now multiply the above equation on the left with $\mathbf{V}_1^{\dagger}$ and on the right with $\mathbf{V}_1$, and noting that $\mathbf{V}_1^{\dagger}\check{\mathbf{G}}\mathbf{V}_1 = \tilde{\boldsymbol{\Sigma}}_{\mathbf{H}}^{-2}$, we get

$$\mathbf{V}_1^{\dagger}(\check{\mathbf{D}}+\mathbf{Z}-\mathbf{Q})\mathbf{V}_1 = \tilde{\boldsymbol{\Sigma}}_{\mathbf{H}}^{-2}.$$

Since $\mathbf{V}_2^{\dagger}\mathbf{V}_1 = 0$, the above equation implies that

$$\check{\mathbf{D}}+\mathbf{Z}-\mathbf{Q} = [\mathbf{V}_1 \; \mathbf{V}_2] \begin{bmatrix} \tilde{\boldsymbol{\Sigma}}_{\mathbf{H}}^{-2} & \mathbf{B} \\ \mathbf{B}^{\dagger} & \mathbf{A} \end{bmatrix} \begin{bmatrix} \mathbf{V}_1^{\dagger} \\ \mathbf{V}_2^{\dagger} \end{bmatrix}$$
$$= \mathbf{V}_1\tilde{\boldsymbol{\Sigma}}_{\mathbf{H}}^{-2}\mathbf{V}_1^{\dagger} + \mathbf{V}_2\mathbf{A}\mathbf{V}_2^{\dagger} + \mathbf{V}_1\mathbf{B}\mathbf{V}_2^{\dagger} + \mathbf{V}_2\mathbf{B}^{\dagger}\mathbf{V}_1^{\dagger}$$

for some Hermitian $(n-m) \times (n-m)$ matrix $\mathbf{A}$ and some $m \times (n-m)$ matrix $\mathbf{B}$. Thus we can write $\mathbf{Q}$ as in (27).

The remaining question is to find $\mathbf{A}$ and $\mathbf{B}$ such that the rank condition in (25) is satisfied. To do this, multiplying equation (27) on the right with $\mathbf{D}\mathbf{V}_2$ and on left with either $\mathbf{V}_2^{\dagger}$ or $\mathbf{V}_1^{\dagger}$, and noting that $\mathbf{V}_2^{\dagger}\mathbf{Z} = \mathbf{V}_2^{\dagger}\check{\mathbf{G}} = 0$, we obtain respectively

$$0 = \mathbf{I}_{n-m} - \mathbf{A}\mathbf{V}_2^{\dagger}\mathbf{D}\mathbf{V}_2 - \mathbf{B}^{\dagger}\mathbf{V}_1^{\dagger}\mathbf{D}\mathbf{V}_2,$$
$$0 = \mathbf{V}_1^{\dagger}(\mathbf{Z}-\check{\mathbf{G}})\mathbf{D}\mathbf{V}_2 - \mathbf{B}\mathbf{V}_2^{\dagger}\mathbf{D}\mathbf{V}_2.$$

Noting that $\mathbf{V}_2^{\dagger}\mathbf{D}\mathbf{V}_2$ is full rank and invertible, we then obtain (29).

## C. Convergence analysis for $n > m$

In this Appendix, we analyze Algorithm 2 for $n > m$ to support the observation that $\check{\mathbf{D}}_i\pi$ decreasing in $i$, where $\pi$ is a diagonal matrix with only 1 and $-1$ on the diagonal such that $\mathrm{diag}(\mathbf{Q}_0)\pi \succcurlyeq \mathbf{P}\pi$.

To show $\check{\mathbf{D}}_{i+1}\pi \preccurlyeq \check{\mathbf{D}}_i\pi$, we use induction. For $i = 0$, from (33), we have

$$\check{\mathbf{D}}_1 = \check{\mathbf{D}}_0 + \mathbf{P} - \mathrm{diag}(\mathbf{Q}_0).$$

Since $\mathrm{diag}(\mathbf{Q}_0)\pi \succcurlyeq \mathbf{P}\pi$, it then follows that $\check{\mathbf{D}}_1\pi \preccurlyeq \check{\mathbf{D}}_0\pi$.

Now suppose that the inequality holds for $i = j$, such that $\check{\mathbf{D}}_{j+1}\pi \preccurlyeq \check{\mathbf{D}}_j\pi$. From (34), we then have

$$\check{\mathbf{D}}_j - \check{\mathbf{D}}_{j+1} = (\mathbf{Q}_j - \mathbf{Q}_{j+1}) - (\mathbf{Z}_j - \mathbf{Z}_{j+1}) + (\mathbf{X}_j - \mathbf{X}_{j+1}).$$

First multiplying both sides on the right with $\pi$ and applying $\check{\mathbf{D}}_{j+1}\pi \preccurlyeq \check{\mathbf{D}}_j\pi$, we obtain

$$(\mathbf{Z}_j - \mathbf{Z}_{j+1})\pi - (\mathbf{X}_j - \mathbf{X}_{j+1})\pi \preccurlyeq (\mathbf{Q}_j - \mathbf{Q}_{j+1})\pi.$$

Now multiplying both sides on the left with $\mathbf{V}_2\mathbf{V}_2^\dagger\mathbf{D}_j \succcurlyeq 0$ and noting that $\mathbf{V}_2^\dagger\mathbf{D}_j\mathbf{Q}_j = 0$, we get

$$\mathbf{V}_2\mathbf{V}_2^\dagger\mathbf{D}_j\left[(\mathbf{Z}_j - \mathbf{Z}_{j+1})\pi - (\mathbf{X}_j - \mathbf{X}_{j+1})\pi\right] \preccurlyeq -\mathbf{V}_2\mathbf{V}_2^\dagger\mathbf{D}_j\mathbf{Q}_{j+1}\pi. \tag{38}$$

Since $\mathbf{V}_2\mathbf{V}_2^\dagger\mathbf{D}_j \succcurlyeq 0$, this can lead to

$$(\mathbf{Z}_j - \mathbf{Z}_{j+1})\pi - (\mathbf{X}_j - \mathbf{X}_{j+1})\pi \preccurlyeq -\mathbf{Q}_{j+1}\pi. \tag{39}$$

Note that since $\mathbf{V}_2\mathbf{V}_2^\dagger\mathbf{D}_j$ is rank deficient, strictly (38) need not to always imply (39). However, we find numerically that the premise of $\mathbf{D}_i\pi$ decreasing always holds either from the initial iteration or after a finite number of iterations, regardless of the initial point $\check{\mathbf{D}}_0$. If it holds after a number of initial iterations, we can reset $\mathbf{Q}_0\pi$. For this reason, we conjecture that (39) always holds true after some iterations.

Next multiplying both sides of (39) on the left with $\mathbf{V}_2\mathbf{V}_2^\dagger\mathbf{D}_{j+1} \succcurlyeq 0$ and noting that $\mathbf{V}_2^\dagger\mathbf{D}_{j+1}\mathbf{Q}_{j+1} = 0$, we get

$$\mathbf{V}_2\mathbf{V}_2^\dagger\mathbf{D}_{j+1}\left[(\mathbf{Z}_j - \mathbf{Z}_{j+1})\pi - (\mathbf{X}_j - \mathbf{X}_{j+1})\pi\right] \preccurlyeq 0,$$

which similarly leads to

$$(\mathbf{Z}_j - \mathbf{Z}_{j+1})\pi - (\mathbf{X}_j - \mathbf{X}_{j+1})\pi \preccurlyeq 0. \tag{40}$$

Now from (33) and (34), we have

$$\check{\mathbf{D}}_{i+1} = \mathbf{P} + \text{diag}(\check{\mathbf{G}}) - \text{diag}(\mathbf{Z}_i - \mathbf{X}_i). \tag{41}$$

Combined with (40), since diagonal of a positive semidefinite matrix is non-negative, we have

$$\check{\mathbf{D}}_{j+2}\pi - \check{\mathbf{D}}_{j+1}\pi = \text{diag}(\mathbf{Z}_j - \mathbf{X}_j)\pi - \text{diag}(\mathbf{Z}_{j+1} - \mathbf{X}_{j+1})\pi \preccurlyeq 0,$$

which completes the induction.